\documentstyle[12pt,twoside,fleqn,espcrc1,epsfig]{article}

\newcommand{\AmS}{{\protect\the\textfont2
  A\kern-.1667em\lower.5ex\hbox{M}\kern-.125emS}}
\newcommand{\be}{\begin{eqnarray}}
\newcommand{\ee}{\end{eqnarray}}
\title{Equation of State, Flow, Fluctuations and $J/\psi$      suppression}

\author{ E.V.~Shuryak\address{State University of New York, 
       Stony Brook, NY 11794, USA}
        \thanks{Based on works done in collaboration with C.M.Hung and D.Teaney.
Supported in part by US DOE.}
        }

\begin{document}
\maketitle

\begin{abstract}
Radial flow observed at AGS/SPS energies is very strong, with collective
velocities of matter reaching about 0.5c for central collisions of the
heaviest ions. The lattice-based Equation of State (EOS) is however 
rather soft, due to the QCD phase transition. We show that both statements are
consistent only if proper kinetic-based
treatment of the  freeze-out is made. In fact  chemical and thermal
freeze-out happen at  quite
different conditions, especially at SPS. Event-by-event fluctuations
can
shed new light on this problem. 
We also propose new model of
 ``anomalous"  $J/\psi$ suppression found
 for PbPb collisions, related it  to prolonged lifetime of dense matter due
the  ``softest point" of the EOS.
\end{abstract}

\section{THE RADIAL FLOW: AN INTRODUCTION}
   One of the major goals of heavy ion physics is to learn the EOS of 
 hot/dense
hadronic matter. 
 Although for central collisions of heavy ions  at the
  SPS   the
 energy density of the order of few $GeV/fm^3$ is 
reached, it remains unknown
when and how  the matter becomes (locally) equilibrated. We also do
not know 
 whether  new phase of matter -
Quark-Gluon Plasma 
(QGP) - is actually produced.
One well-known strategy addressing these issues
 relies on
(very  rare) processes happening at earlier
   stages,  the e/m probes \cite{shu_78_photo} or  $J/\psi$
   suppression (see section 5). Both  lead to exciting
 experimental  findings, much discussed at this conference.

Another approach (to be mostly discussed in this talk)
  is based on hadronic observables,  well measured now in
  high-statistics
experiments. Although
re-scatterings  tend to erase most  traces of the
   dense stage, some of them     are preserved and
 $accumulated$ during the expansion:
  {\it collective  flow} is one of them. Existing data 
strongly suggest that the hadronic system  does indeed behave
as
a truly macroscopic one. Rather detailed phenomenology of the so
called $directed$ flow was covered here by  Ollitrault, so
I focus on $radial$ (axially symmetric)
flow\footnote{ The $longitudinal$ flow was discussed
in multiple hydro-based works. However, due to uncertainty in
initial conditions,  predictive power
of hydrodynamics is rather limited.
} for  central collisions. 

  Its very existence  was widely debated for years, but
(although at QM97 we have still witnessed remnants of this debate)   
it seems to be proven now ``beyond a reasonable doubt''. 
The HBT data, Coulomb effects (or any reasonable event generator)
show that transverse size at freeze-out  significantly exceed that of parent
 nuclei.   An excellent
test for existence of the flow 
is provided by deuterons.
 The shape of their  spectrum, its  slope $\tilde T_d$
and even $yield$ are all very sensitive to it. The
 flow implies  a  
correlation between position and momentum, which helps to produce
deuterons.
If this correlation is
 removed (see \cite{deutron_RQMD}, where
 in RQMD output the nucleon's positions or momenta were interchanged) 
 deuteron spectra dramatically change.

  Phenomenology of the $m_t$ slopes\footnote{The
 slopes are not temperatures,
 as they also include effect of the flow to be discussed,
 and resonance decays. }, 
 $\tilde T$, is  qualitatively explained
by flow.
 However, looking at it quantitatively one finds that their
 understanding 
 should be improved. True,
  one may get a very good fit with 
 some velocity profiles  and  $fixed$  decoupling
temperature
 $T_f$.
But the  spectra 
 allow for multiple fits, with wide  trade-offs between  
 $<v_t>$ and $T_f$. Furthermore, those fits are based on grossly oversimplified
picture: (i) the ad hoc profiles do not correspond to hydro with a
realistic EOS; (ii) 
the main assumption -  for all secondaries
one should  expect the same  $v_t$ and $T_f$ - 
 obviously contradicts to
elementary kinetics. Different secondaries do have very different
cross sections, and therefore  decouple from flow at different times.

  As one can see from experimental talks at this conference, 
by now we have rich experimental
    systematics of $m_t$ slopes.
First of all (as shown  by NA44 at QM96), the slopes have very
strong A dependence. While the pp data show no
 trace of the radial flow  (as noted long ago \cite{SZ}), it is seen
 for nuclear collisions. Furthermore,  $<v_t>$ is
 increased by about factor two from SS to PbPb.
Such trend qualitatively contradicts to predictions of
 most of the  hydro models in literature, with
 A-independent freeze-out.
 
Then there is significant dependence on particle species:
while  $\pi,K,N,d$ slopes show about linear increase with the particle
mass.
 New data for strange hadrons $\phi,\Lambda,\Xi,\Omega$ deviate
from this line. (This does not  mean that they do not flow
together with others for some time - they may be just decoupled
 earlier, due to smaller cross sections.)  

  The next point: data show  strong $rapidity$ dependence
of the radial flow. 
The strongest effect (and its A-dependence)
 comes preferentially from mid-rapidities.

  Finally:   dependence   on
the {\it collision energy} E/A. For 
 Bevalac/SIS energies $<v_t>$
steadily grows with E/A, and new low energy runs at AGS show that it is also
the case in the whole AGS domain. At
11 GeV/A flow  velocity is about the same as at SPS (160-200 GeV/A).
It is very important to know what happens in between: 
most probably there exist  a
maximum at some energy\footnote{Unless
a
saturation occurs exactly
 10 GeV/c. But as EOS is  changing drastically for
corresponding energy densities (both due to
changing meson/baryon ratio and   to the phase
transition), exact cancellation of its influence on flow is unlikely.}.

\section{THE EQUATION OF STATE}
\begin{figure}[t]
\includegraphics[width=6.3cm,angle=-90]{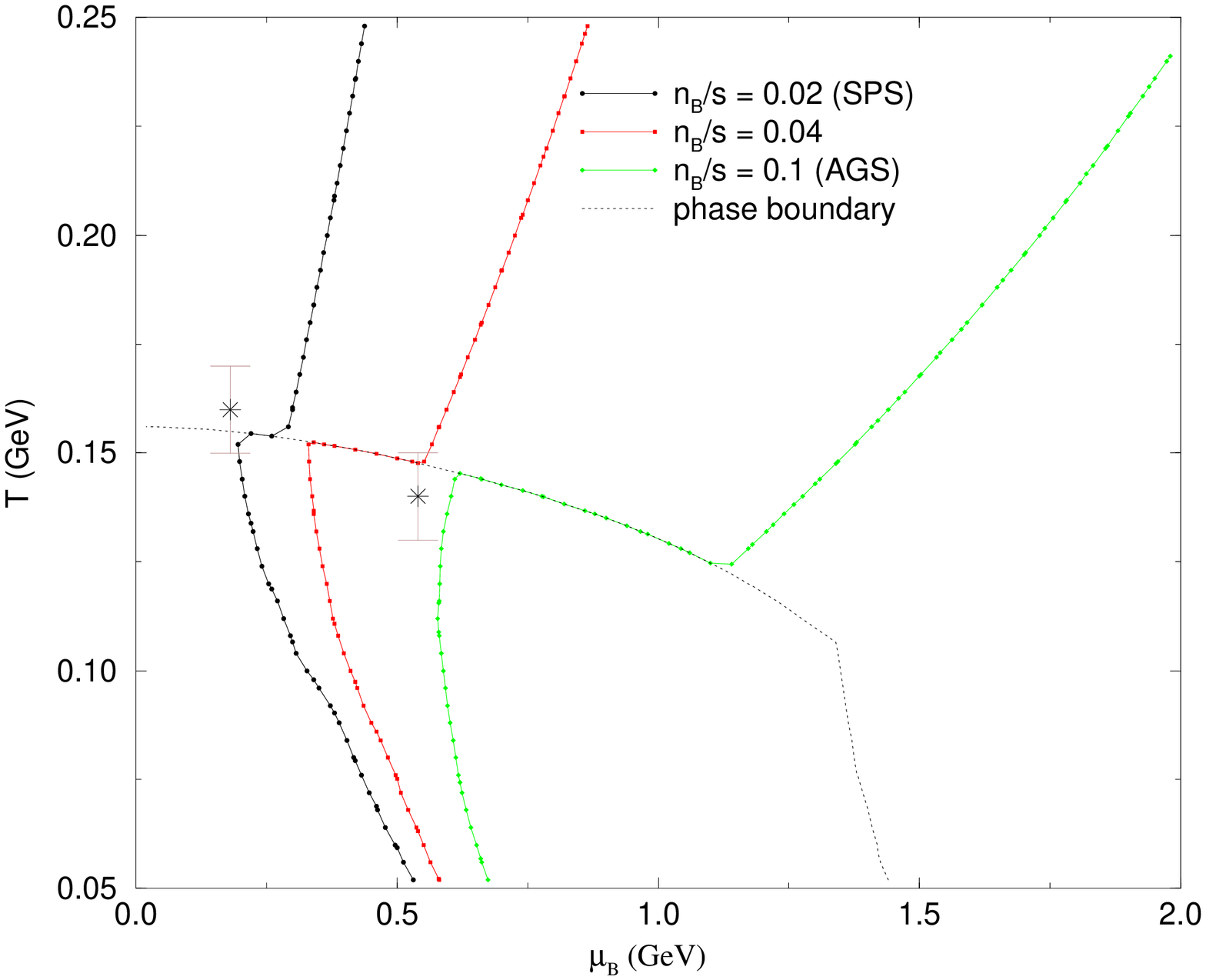}
\hspace{-.4cm}
\includegraphics[width=6.3cm,angle=-90]{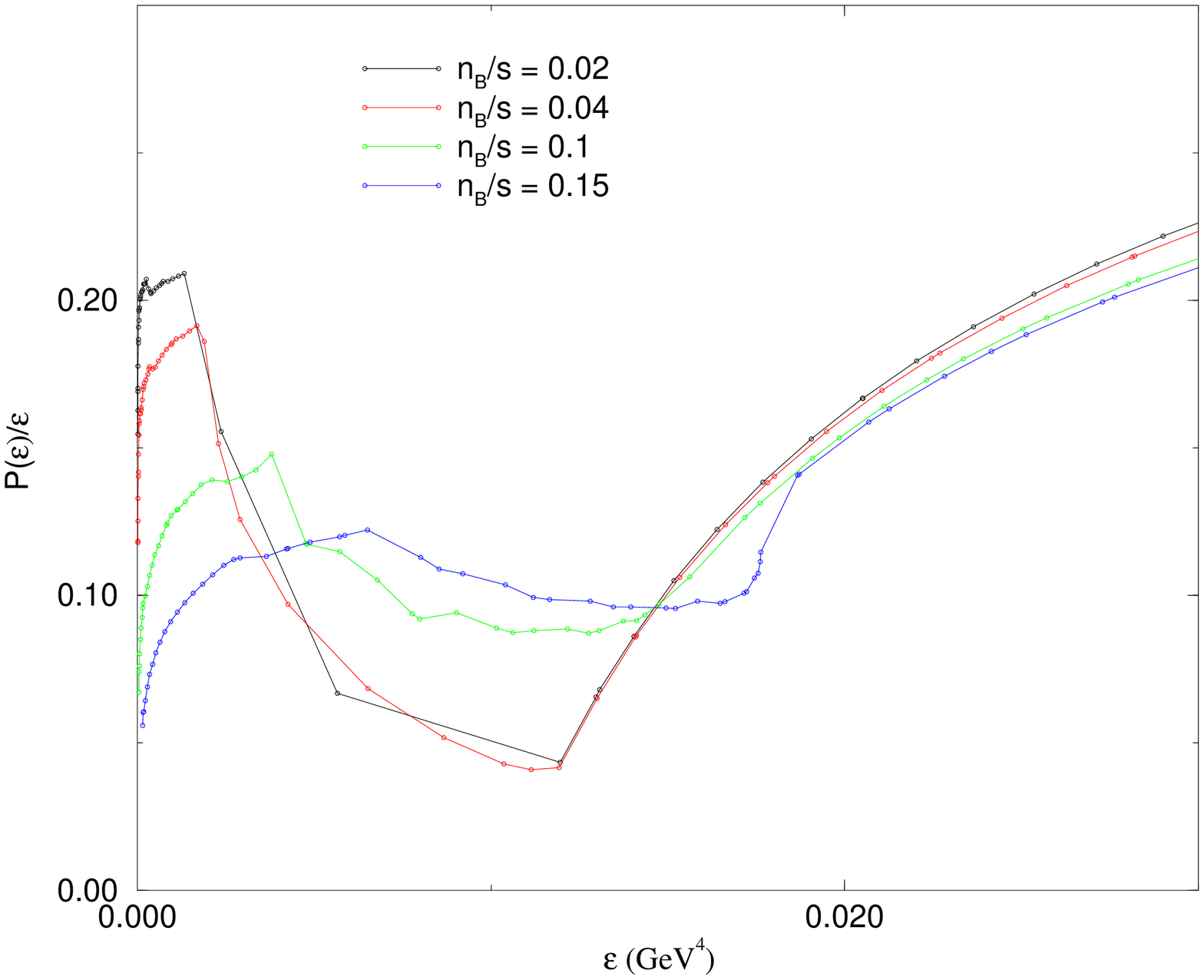}
\caption{\label{fig_EOS}
 (a) Paths in the $T-\mu$ plane for different
baryon admixture, for resonance gas plus the QGP; 
(b) the ratio of pressure to energy density $p/\epsilon$ versus
$\epsilon$,
for different baryon admixture. }
\end{figure}

   We use rather standard EOS of hadronic matter, 
 a {\it resonance gas} for hadronic phase \cite{resonancegas} and 
a simple bag-type quark-gluon plasma,
 with a Bag constant fitted to $T_c= 160 MeV$.
  	In Fig.\ref{fig_EOS}(a) we show its phase
        boundary\footnote{This model is known to have wrong behavior
 of the phase
          boundary at large densities, and even with excluded volume
          for baryons (which we use) one finds some problems.
Significant progress in understanding of high-density matter was made recently,
see talks of T.Schaefer and K.Rajagopal.
} and the
 paths the matter elements follow during
$adiabatic$ expansion.
 As both baryon number and entropy is conserved, the lines are
marked by their ratio. Those for
$n_b/s=0.02,0.1$ correspond approximately to SPS (160 GeV A) and AGS (11 GeV A) heavy ion
collisions,
respectively. (Note  a non-trivial zigzag shape,
with slight re-heating in the mixed phase.) The paths end in hadronic phase by
freeze-out: as we discuss below, it is non-universal
for different collisions and even volume elements.
Two points in   Fig.\ref{fig_EOS}(a) correspond to chemical
freeze-out, extracted from
thermal fit to  particle composition observed at AGS and SPS
(from \cite{PBM_etal}). 
 Both (inside error bars) coincide with the left corners
of the zigzag  path. 

  The final velocity of the observed collective ``flow''
is time integral of the acceleration, which is proportional to $p/ \epsilon$
ratio plotted in Fig.\ref{fig_EOS}(b).
Note that  the QCD resonance gas 
in fact  has  a very simple EOS
$p/ \epsilon\approx const$, while in the ``mixed phase" is  softer indeed.
Furthermore, there is a deep minimum known as the ``softest
point'' \cite{HS_96}.
The contrast between ``softness" of matter at dense stages and relative 
``stiffness"  at the dilute ones is much stronger for the SPS case,
because there are less baryons. 

\begin{figure}[t]
\includegraphics[width=12cm]{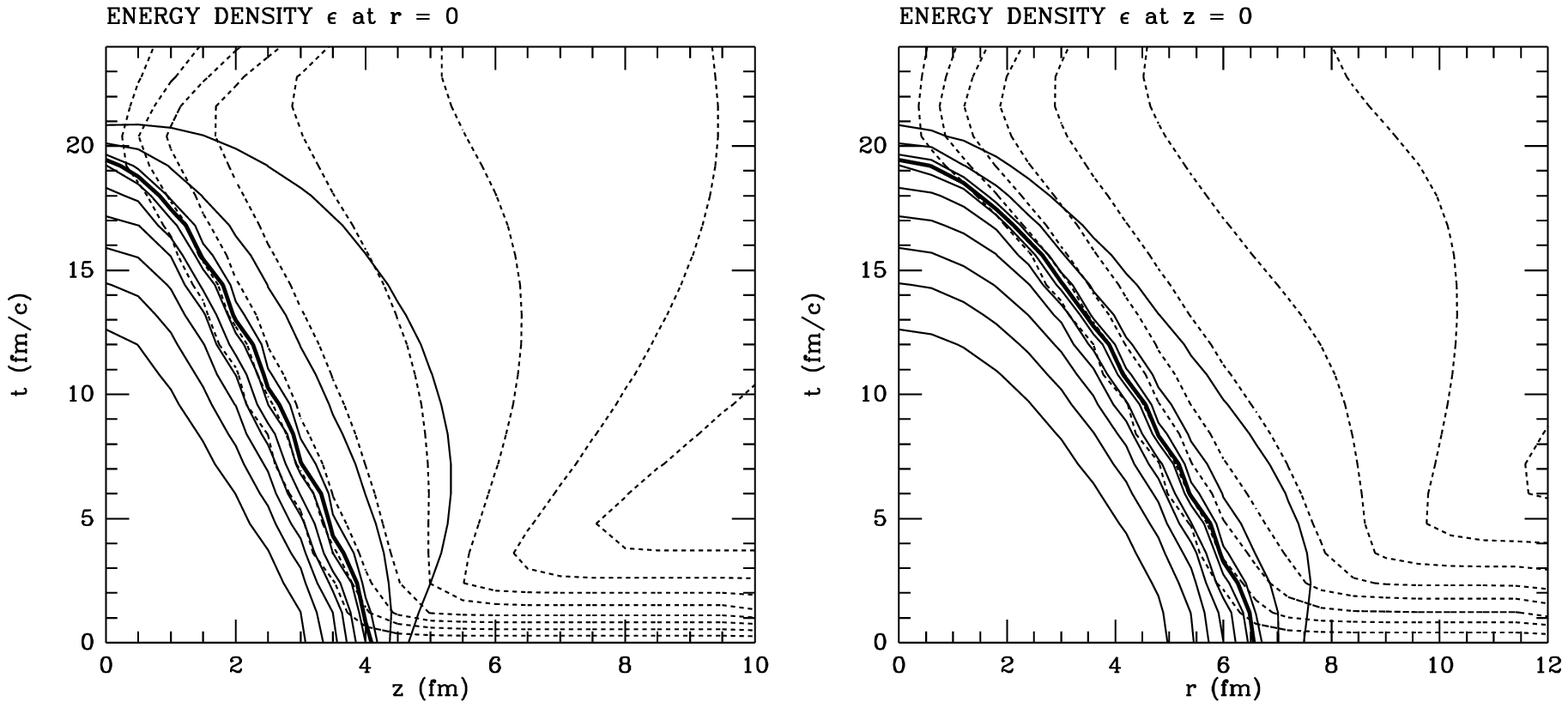}
\includegraphics[width=12cm]{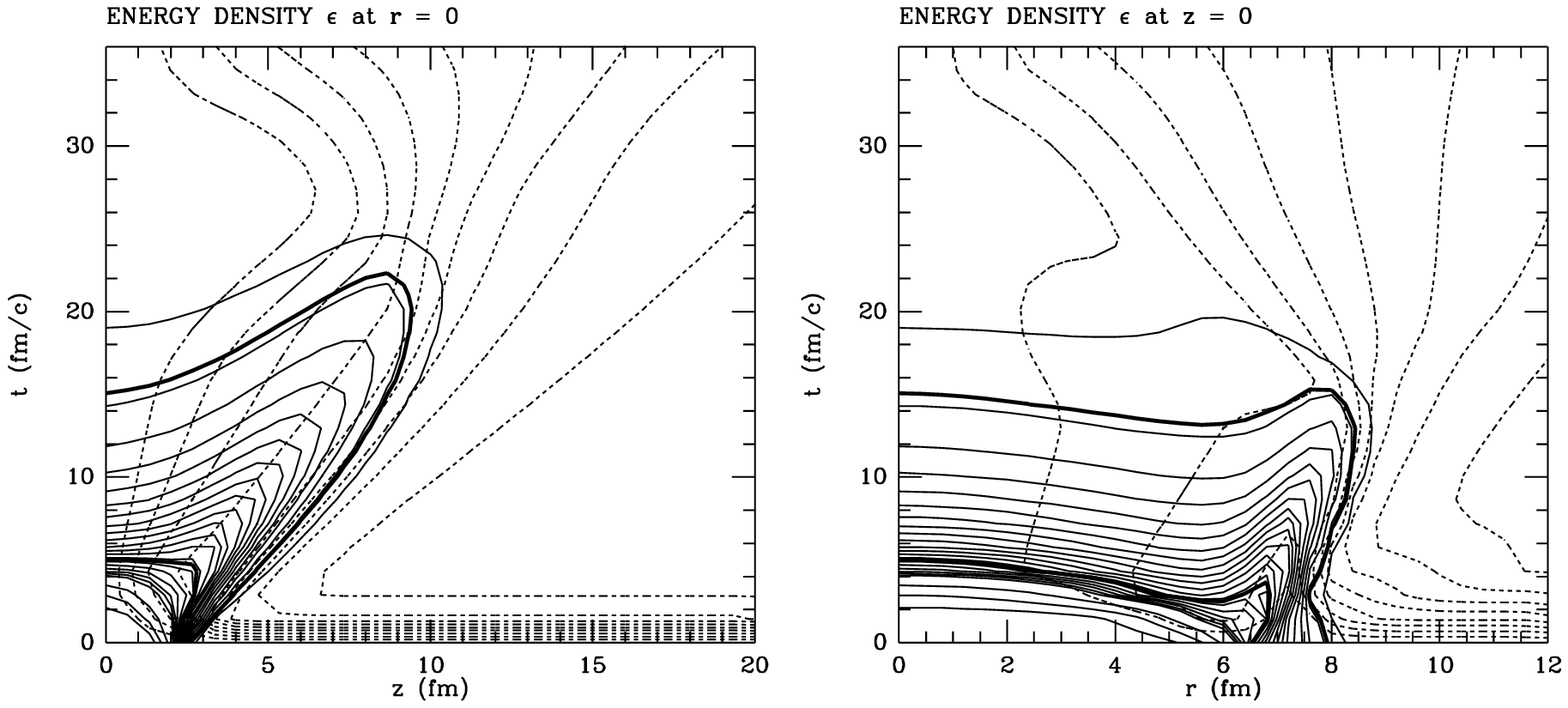}
\caption{\label{fig-hydro}
 Hydrodynamical solutions for central
 (a) 11.6A GeV Au+Au and (b) 160A GeV Pb+Pb. The solid contours are
energy density contours, with the bold contour being the
boundary between the mixed and hadronic phase ($\epsilon = 0.35 {\rm
  GeV/fm^3}$). The dotted contours are the longitudinal (left) and
radial (right) velocity contours, with values starting from the left
of 0.01, 0.05, 0.1, 0.2,...}
\end{figure}

Typical solution for 11.6A GeV Au+Au is shown in
Fig.\ref{fig-hydro}
while for 160A GeV Pb+Pb it is shown in Fig.\ref{fig-hydro}.
 First of all, they are
qualitatively different. At AGS  the
longitudinal and transverse expansion are not qualitatively
 different, while at SPS
 the $longitudinal$ flow has already distinct ultra-relativistic
 features,  most isotherms being close to Bjorken
hyperbola $\tau=\sqrt(t^2-z^2)$. Less
obvious observation
(resulting from particular EOS with the QCD phase transition)
is  a dramatic difference in the $transverse$ expansion.
 The AGS case can be described as ``burning in'', the lines of
constant
energy density moves inward with some small constant speed. At SPS
the mixed phase matter burns into the low density hadron gas at a
cylinder (known also as a
``burning log''), which has  nearly time-independent 
 transverse radius 6-8 fm.

\section{KINETICS OF THE FREEZE-OUT}

 We have already remarked that in order to explain flow (and many
 other things)
 quantitatively, the
hydro solution should be complimented with (well known!) 
  kinetics of the dilute hadron matter at the end. So the 
    $practical$ objective of \cite{HS_97}  was to create
a {\it next generation} Hydro-Kinetic model
for heavy ion collisions,  
HKM for short. It incorporates
basic elements of the macroscopic approach --
(i) thermodynamics of hadronic matter, (ii)  
hydrodynamics of its expansion, and (iii)
realistic hadronic kinetics at the
freeze-out, plus practically important (iv) resonance decays. 
Most elements of the model have in fact been worked out in literature,
 some are new, but  they are practically taken together  for the first
 time.

  Before we go into specifics, let us make brief comments on 
literature. Many
  hydro-based works have  
parameterized  the initial conditions in a way which
results in  rough reproduction of the
 $y,p_t$ spectra of different species, but none so far addressed the
 details of the radial flow systematics discussed above.
 Probably the
closest in spirit to our work is recent paper \cite{MH} \footnote{ 
  This paper  
also uses local freeze-out conditions.
Unfortunately, their method (referred to as ``global'' hydrodynamics) include
transverse plane averaging, which is unnecessary
and  significantly obscures the results obtained. 
However, to the extent flow is concerned, 
 our findings  agree.}

  The
  cascade ``event generators'' (Fritjof,Venus,RQMD,ARC
etc) are widely used, and  (at least) RQMD 
provides a reasonable  radial flow. (They are  much
more beloved by experimentalists than theorists, who of course know that
all of thesem models know
nothing about QGP, and thus
 {\it directly contradict} to QCD theory/lattice.)   
Hydrodynamics
 and cascades are not physics alternatives, but
 just  complementary tools,  most useful for dense and    
dilute conditions, respectively.
These days, 
 with approaching  RHIC/LHC, one has  to
work with thousands of secondaries.
Direct simulation of all  re-scatterings are neither practical not
necessary:
as soon as the system
becomes  much larger than the correlation length, it
can be cut into independent
``cells''. The most reasonable strategy for  RHIC/LHC is, to my mind,
a combination (parton cascade)-(hydro)-(hadronic cascade).

 One should make clear
 distinction between $chemical$ and $thermal$ freeze-outs.
Reactions changing particle composition 
have cross sections very different from those for  elastic
collisions. 
 In most hydro-based papers published so far it was however ignored, and
  the expansion was simply cut off
at fixed T, usually about 140 MeV. It is clear now that at SPS
chemical
freeze-out happens at $T_{ch}\approx 160$ MeV, while thermal one occurs (for
heavy ions, in their center) to temperatures as low as $T_{th}=100-120 MeV$.
It does not look like a huge
difference,  but in terms of density, space-time picture and flow
it is is crucial. Extra few fm/c
time available at the end for  heavy
ion collisions (PbPb relative to SS)
leads to  ``extra push" by stiff hadronic gas, especially for nucleons.
(This is where the twice stronger collective flow for heavy ions
comes from.)

  Apart of flow, there are of course other means to verify the
  difference
between 
   chemical and thermal freeze-outs. If they do not coincide, the 
  chemical potentials cannot stay at zero\footnote{It was originally pointed
out  by G.Baym, see  quantitative discussion in   \cite{Bebie_etal}.}.
The effect should in fact be directly observable in $p_t$ distribution of 
 pions\footnote{For
  clarity: those potentials are conjugated to  
total number of particles, so say for pions they enter distributions of
$\pi+,\pi^-,\pi^0$  with the same sign. } as
 extra low $p_t$ enhancement (on top of resonance decays). 
We have calculated that for  central PbPb at SPS  it should reach
$\mu_\pi=60-80 MeV$. Taking the ratio of NA44 data for PbPb/SS
we  indeed found
extra low-$p_t$ enhancement compatible with it.


\begin{figure}[h]
\begin{center}
\includegraphics[width=11.cm]{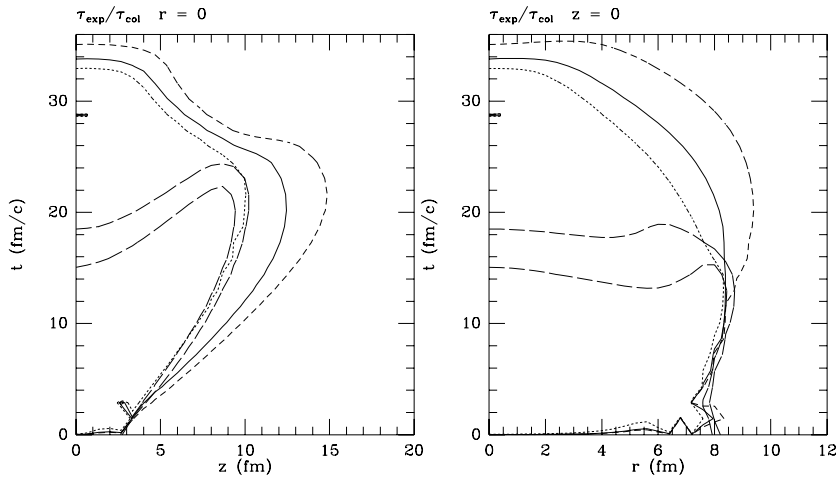}

\caption{\label{fos} Freeze-out surfaces for
 160A GeV Pb+Pb collisions. Solid, short-dashed and dotted curved are
for pions, nucleons and kaons, respectively. Two long-dahsed curves
are isoterms with $T=140$ (upper) and 160 MeV (lower).}\end{center}
\end{figure}

   HKM  uses  ``local''  kinetic condition\cite{BGZ_78}:
 the 
(relevant) collision rate $1/\tau_{coll}$ becomes comparable to 
 the expansion rate $1/\tau_{expansion}=\partial_\mu u_\mu$.
Using it in a form  $\tau_{expansion}/\tau_{coll} =1/2$  we  determine
the freeze-out (3-d) surface. Representative case is shown
in Figs.\ref{fos}, as  sections
by time t - longitudinal coordinate z plane (at transverse
coordinate r=0) and the t-r plane (z=0).
Note significant dependence on the particle kind.
The larger is the system, the lower is $T_{th}$ on this
surface.
Furthermore, the shape of the freeze-out surface is very different
from that of isotherms. It means that
there is a significant variation of this temperature
over the surface itself:
in order to find the coolest pion gas, one should
look at the very center of
central collisions of heaviest nuclei at highest energy!

\section{THE RADIAL FLOW: SOME RESULTS}
 The improved 
  freeze-out condition leads to huge 
difference   for the flow.
Although it does not significantly prolong total
lifetime, it significantly
increases  the lifetime of hadronic phase at which the matter is 
most ``stiff''. 

\begin{figure}[ht]
\includegraphics[width=8.cm]{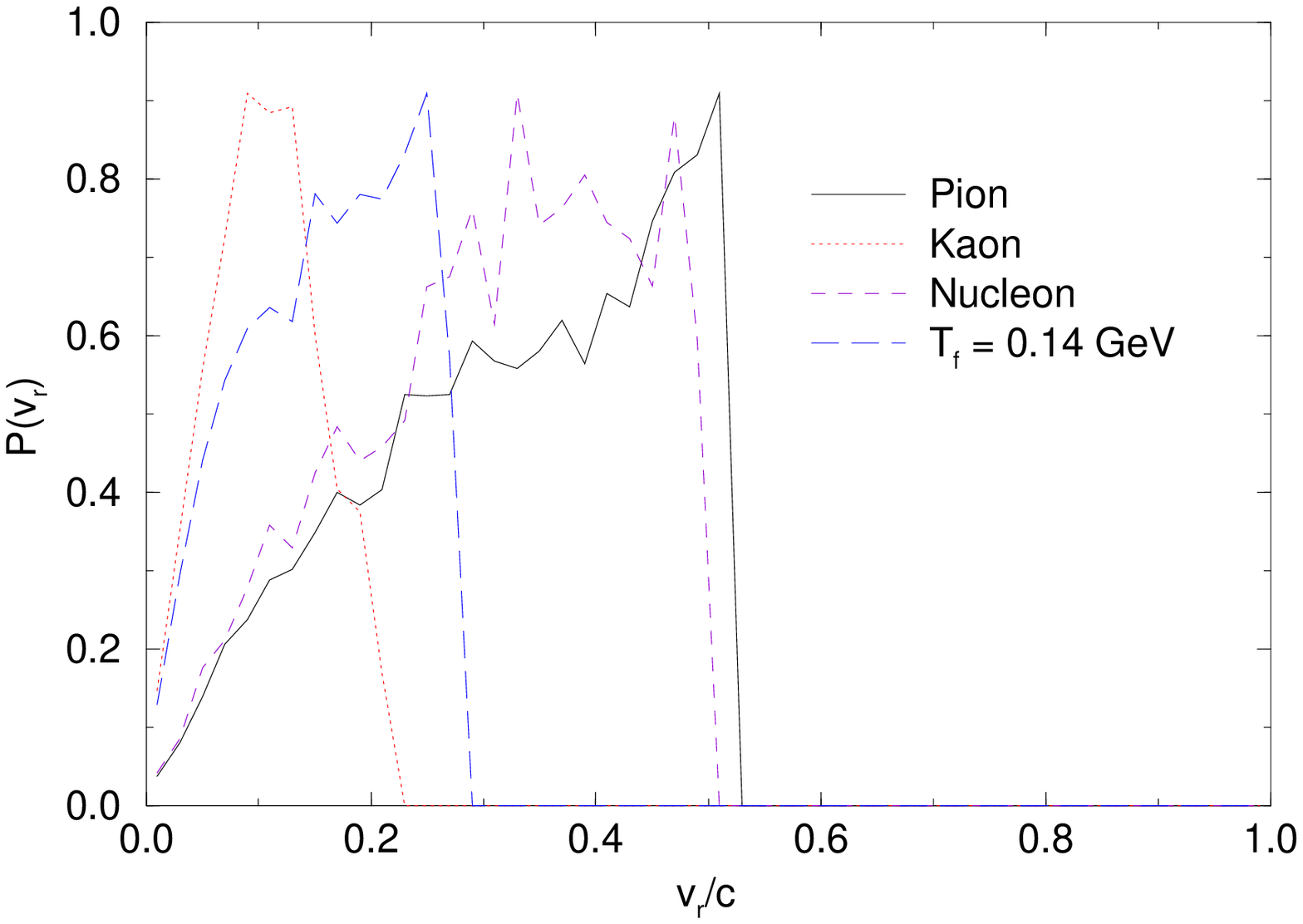}
\hspace{-.2cm}
\includegraphics[width=8.cm]{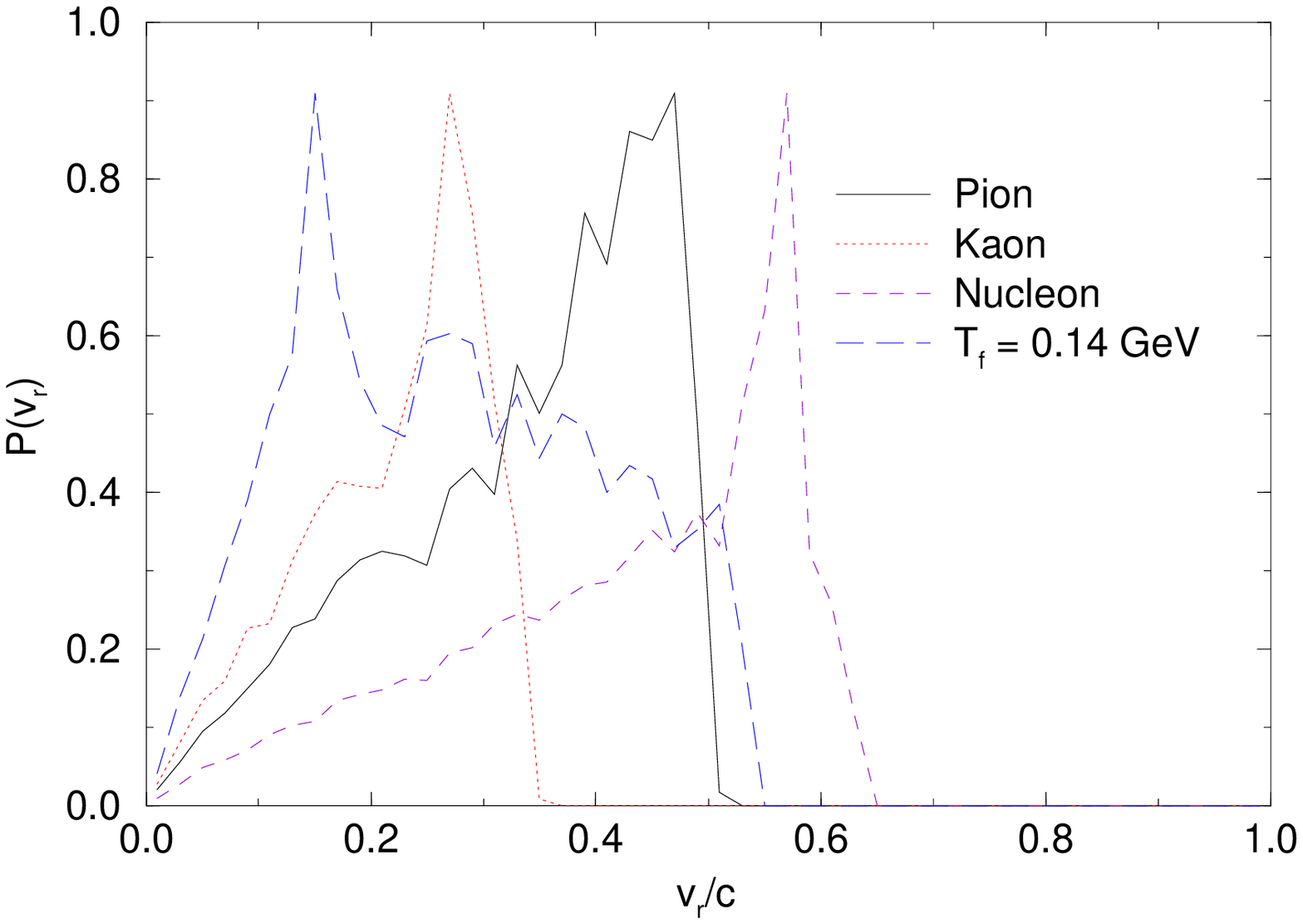}
\vspace{-.3cm}
\caption{\label{out213vr} Transverse velocity distribution over the
  various freeze-out surfaces for11.6A GeV Au+Au collision (a)
and  160A GeV Pb+Pb (b).}
\end{figure}

\begin{figure}[h]
\includegraphics[width=8.cm]{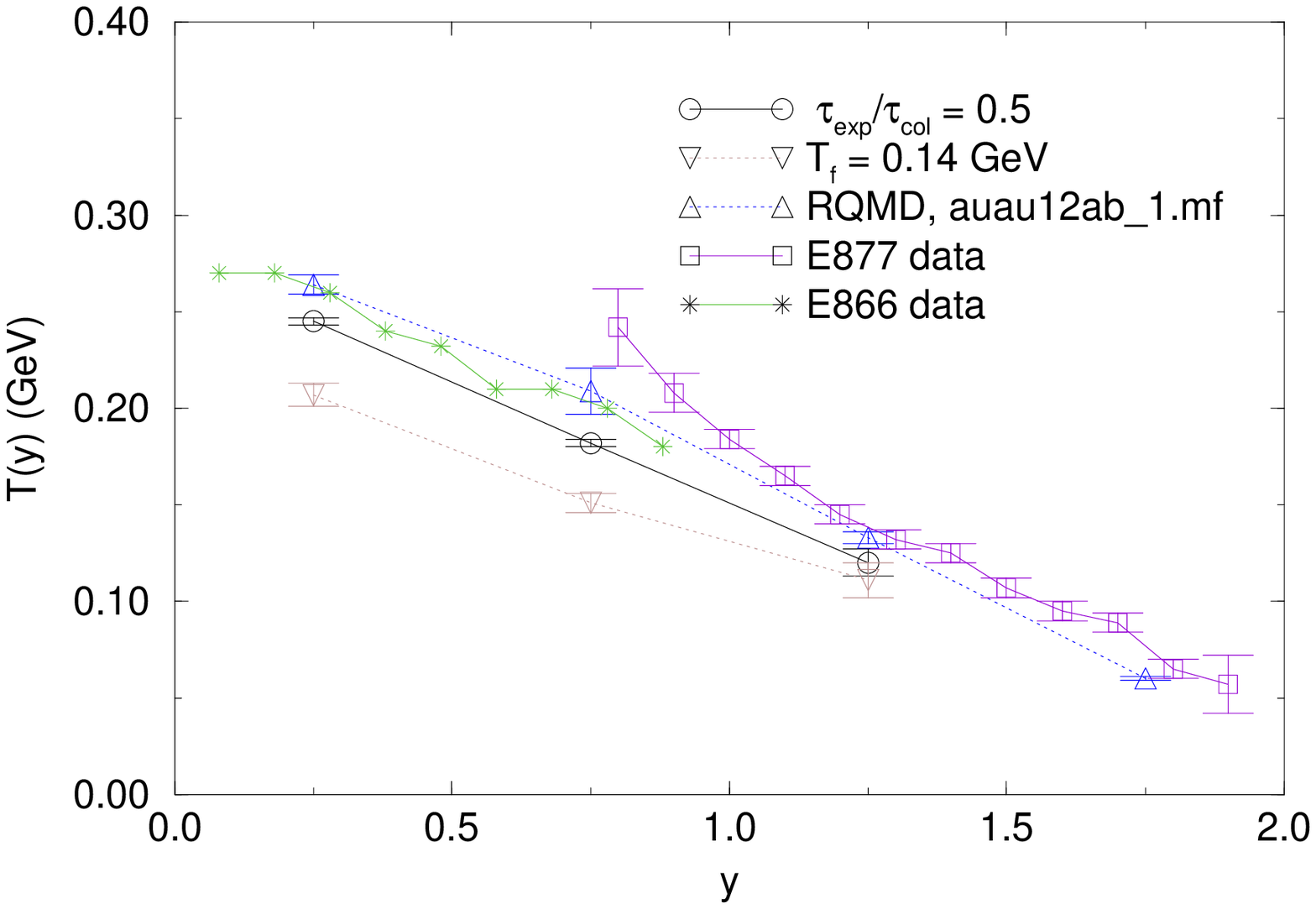}
\hspace{-.2cm}
\includegraphics[width=8.cm]{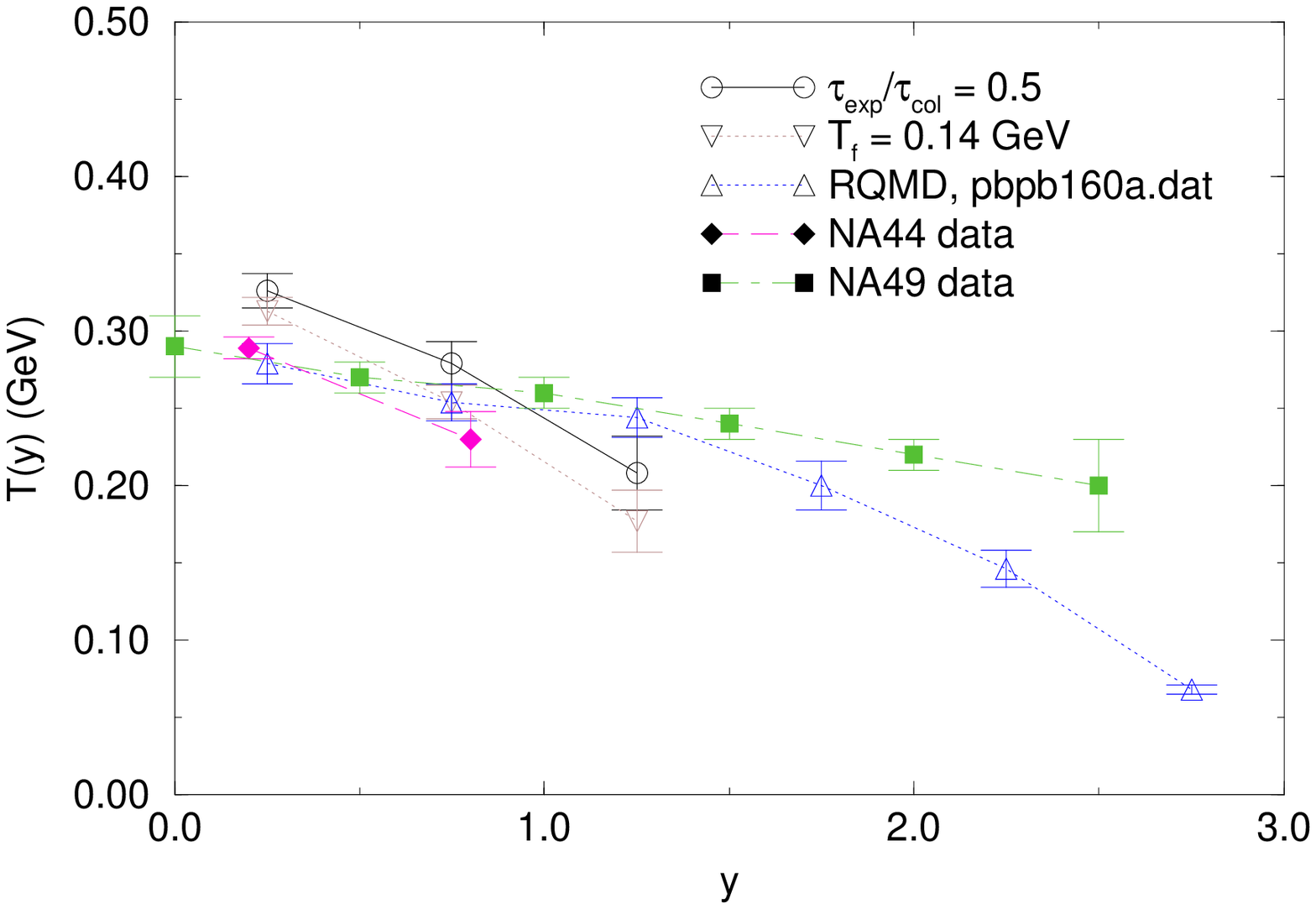}
\caption{\label{flowslopes1} Nucleon slope parameters for 11.6A GeV
  Au+Au (a) and  158A GeV Pb+Pb (b) }
\end{figure}

Distribution over transverse velocities calculated at the freeze-out surfaces
 is shown in Figs.\ref{out213vr}.
These distributions
have sharp peak at their right end,  more pronounced at SPS.
Its position depends significantly on the particle type, reaching
 $v_t=0.6$
for N at SPS. (However
velocity distribution for the isotherms $T=.14 GeV$
  peak is at
  much smaller  $v_t\approx .17$.
This difference is smaller for medium ions, not shown.)
In Figs.\ref{flowslopes1} 
 we show how this translates into
 the observable quantity\footnote{If we would plot just $p_t$ spectra,
   one would see no difference.  }, the   $m_t$
 slopes    $\tilde T(y)$ plotted 
as a function of rapidity y.
  For AuAu data at AGS  Fig.\ref{flowslopes1}(a) one can see, that
our results are slightly below RQMD ones:
it is because this version of RQMD has
 a (specially tuned)
repulsive baryon-induced potential, on the top of the pure cascade.
 Approximate agreement with RQMD
 at SPS energies is due to ``softness'' of
  RQMD EOS:  at early times the energy is stored not  in
 resonances, but
  in (longitudinally
 stretched) classical strings, producing no pressure in
 the transverse direction.

 Let us  stress that we have not attempted any fine tuning of the
  parameters. The main ingredients, EOS and freeze-out condition 
 were fixed at an early stage and not modified when hadron
spectra/slopes
were calculated. 
Our main objective was to test mostly the systematics of
the flow discussed in the introduction.

\begin{figure}[h]
\begin{center}
\includegraphics[width=5.2cm]{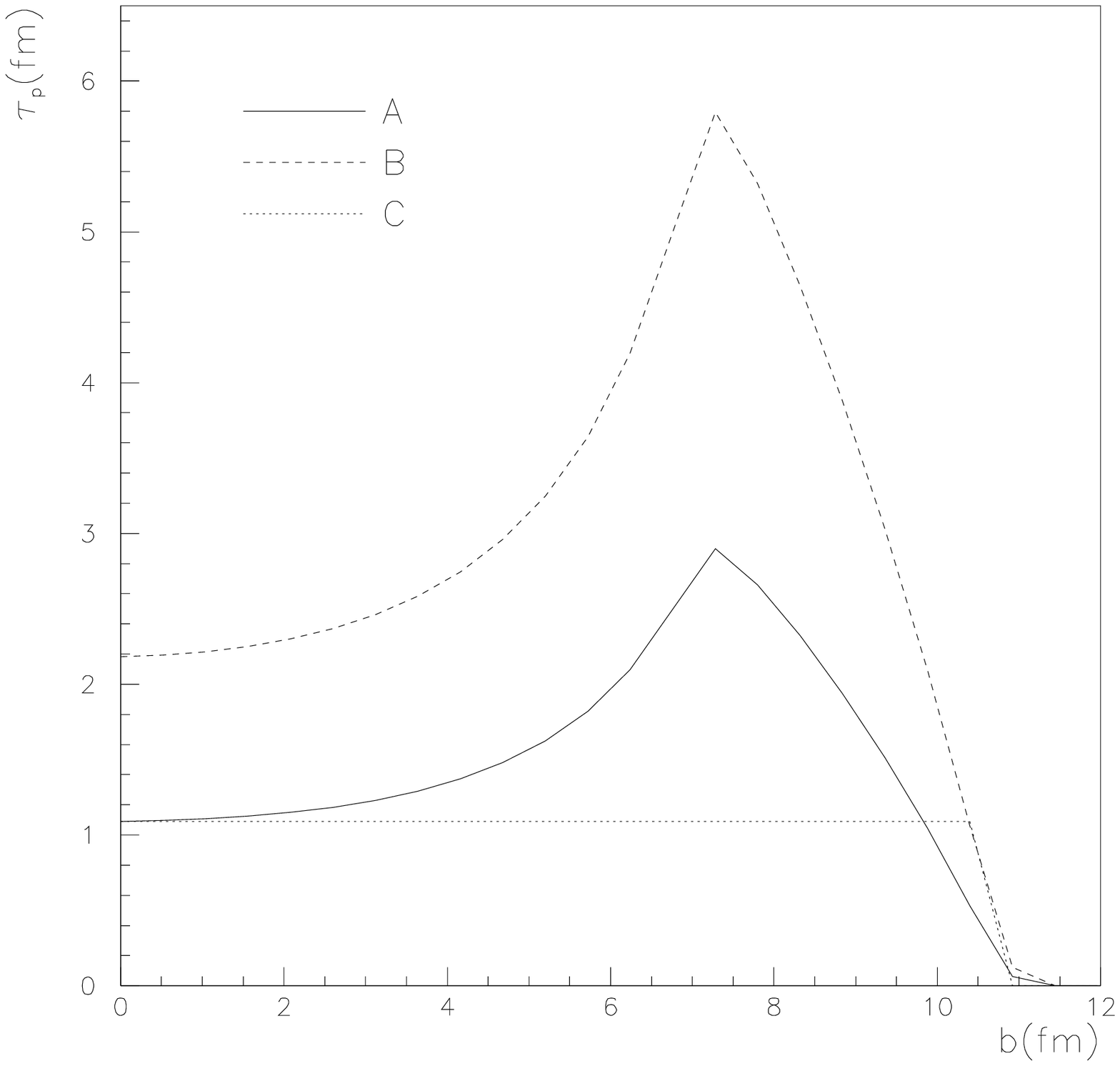}
\hspace{-.3cm}
\includegraphics[width=5.2cm]{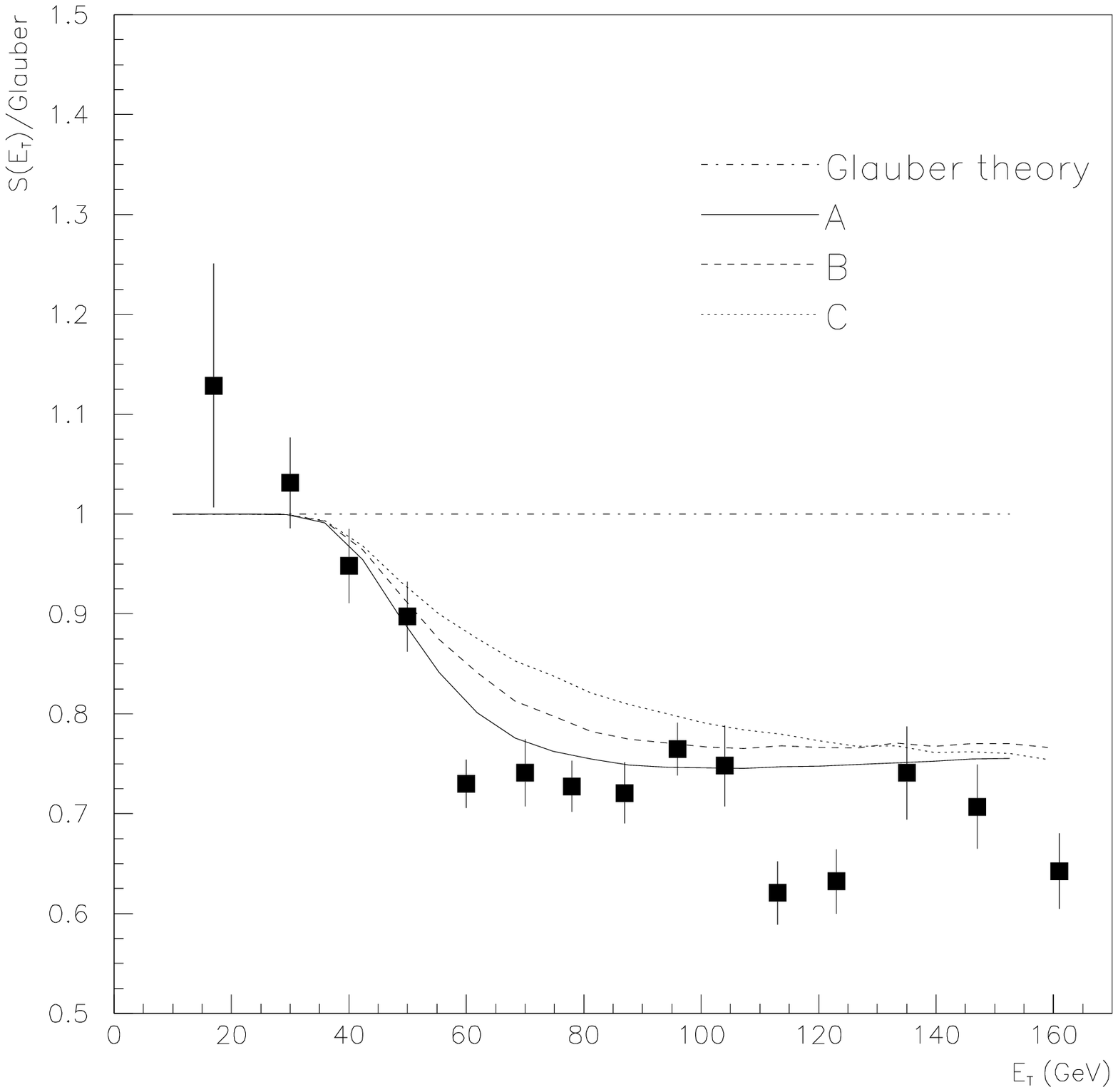}
\hspace{-.3cm}
\includegraphics[width=5.2cm]{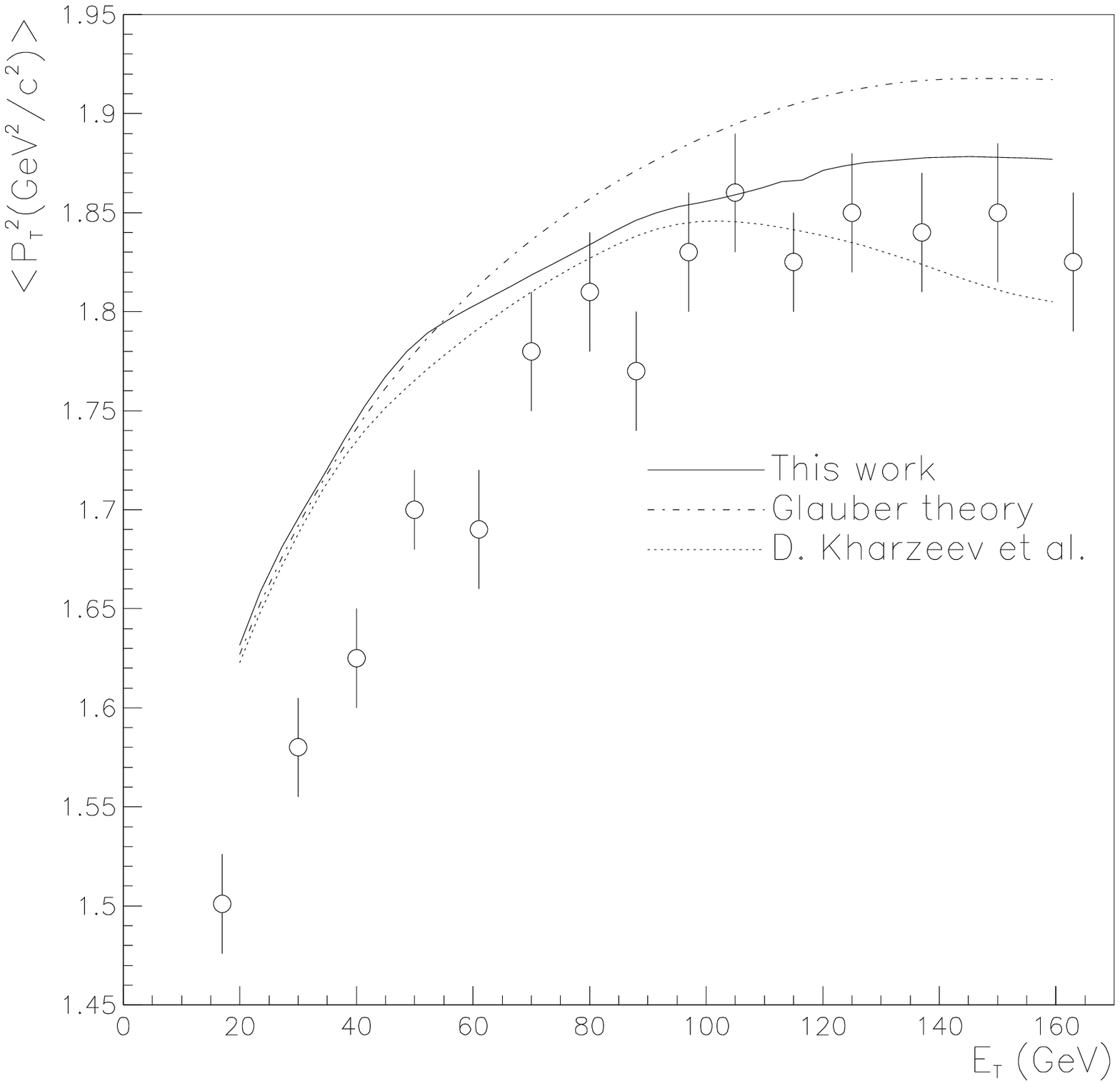}
\end{center}
\caption{\label{suppression}
(a) Three variants of QGP lifetime as a function of impact parameter b lead 
to  $J/\psi$ suppression (relative to Glauber theory) shown in
(b). 
The points are preliminary NA50 data. (c)The $J/\psi$ mean $p_T^2$
versus
treansverse energy $E_t$ . ``This work'' means variant B in figs (a,b).}
\end{figure}

\section{THE ``SOFTEST POINT" AND THE $J/\psi$ SUPPRESSION}
 Recent studies by the NA50 collaboration \cite{na50}
 of the $J/\psi $ suppression in
Pb(158 GeV-A)-Pb collisions
 show   dramatic
differences with
the extrapolations of known trends, from lighter beams and pA  
collisions. These data, if correct, imply existence of a new suppression
mechanism, which sets in\footnote{ The surprising and much debated
 feature of the data
is that this new regime seems to set in extremely sharply, see
fig.\ref{suppression}. Let me therefore stress again, 
 that we try to describe  trend in the whole region, not just its onset.} for PbPb for 
impact parameters   $b < 8 fm$.) 
They
 may be a long-awaited manifestations of  
  collective effects, related to dramatic
changes in properties
of hot/dense hadronic matter related to the QCD phase transition.
In general, ``anomalous"  $J/\Psi $ suppression
can be caused either by (i) increased absorption rate or
(ii) by increased time the  $J/\Psi $ spends in a dense/hot matter. 
The available literature is mostly devoted to the former possibility,
while  we study  the latter one. 

 The oldest (and the simplest idea \cite{shu_78_photo}) is that if Quark-Gluon Plasma (QGP)
  is formed, its free gluons  can rather easily
 ``ionize'' charmonium states by photo-effect-type
reaction   $gJ/\psi\rightarrow \bar c c$ 
 \cite{peskin,KhS_photo,KNS}. Another
 idea  \cite{ms,KS} is that  successive charmonium levels
$\psi',\chi,J/\psi$ should
"melt" (fail to exist) at increasing density
 due to Debye screening in the Quark-Gluon Plasma (QGP).

 There is a significant  difference between
these two mechanisms.  ``Melting'' scenario  
implies existence of certain $thresholds$ as a function of energy
density, and one can indeed expect rapid onset of the
suppression.
Furthermore, it naturally explains why different charmonium states 
get suppressed at different $\epsilon$ (b, $E_t$). Although the
realistic model of this type is yet to be worked out,
 this scenario may be potentially successful. 

In view of NA50 data, scenario 
 based on the  gluonic ``photo-effect" has problems. 
Like any other probabilistic process   
(``co-mover'' absorption) it
cannot create discontinuous behavior by itself.
  Therefore 
an additional hypothesis was made \cite{KNS} -  it is
the {\it formation} of QGP which is discontinuous, it happens
suddenly after a certain critical energy
density $\epsilon_c$ is reached.
But
 thermodynamics tells us that in a first-order
 phase transition  there is discontinuity in $temperature$ but not in
{\it energy density}: the  mixed phase interpolates linearly between
the energy densities of the two phases,  $\epsilon_{min}$ and  $\epsilon_{max}$.  
Relying on non-equilibrium fluctuations require too much fine tuning.
Furthermore, this scenario has a phenomenological problem:
 any sudden transition  into QGP phase at some $\epsilon_c$  results
 in a jump of the   total entropy produced, while 
the data on  
$<n_{ch}(E_t)>$ does not seem to have a jump at $E_t\approx 50 GeV$.  

 We have mentioned above that 
QCD phase transition leads to a very non-trivial EOS, with
  a deep minimum in $p/\epsilon$ at the
``softest point'', $\epsilon_{max}\sim 1-2 GeV/fm^3$.
 It was pointed out in \cite{HS_96}  
(see also later study \cite{Rischke}) that  hydrodynamical expansion
 with such EOS has a non-monotonous dependence on initial conditions.    
When initial energy density is close to the ``softest point'',
  the expansion is more slow. As a result,  the QGP lifetime has
 a peak as a function of collision energy (or impact parameter b),
increasing by significant factor (2-3),
 as compared to that at central collisions at
SPS.

 In \cite{ST_97} we study whether the non-trivial 
features of $J/\psi$ suppression can be explained by such increase
in the QGP lifetime.
Before we go into specifics, let us point out qualitative differences
between this scenario and the one based on $J/\psi$ ``melting''.
The main difference is that in the latter case the complete
suppression of corresponding  charmonium states
are assumed for $all$ $\epsilon>\epsilon_c$, while in our
case the anomaly is concentrated around $\epsilon\approx
\epsilon_{max}$ and at larger $\epsilon$ the survival probability
$increases$ back. 

Estimating initial dependence of energy density
(e.g. by ``wounded nucleon model''), one
can see that indeed the impact parameter $b\approx 8 fm$ in PbPb collision
roughly corresponds to the ``softest point''.
The
hydro expansion for non-central collisions is  complicated 
by the so called directed flow. Fortunately for    $J/\psi$ suppression 
 one may ignore it: expansion is predominantly longitudinal  during the
 first few fermi/c.

 So, assuming that absorption rates are unchanged, we study
whether changes in QGP lifetime can 
affect $J/\psi$ suppression and reproduce NA50
data. The main problem here 
 is ``leakage'' of $J/\psi$ from the system, which  limits the
sensitivity of $J/\psi$ yield to the QGP lifetime.
We have reproduce  known observables (distribution in
transverse energy $E_t$ and its correlation with b)
by a small Monte Carlo model. For  Pb(158-GeVA)-Pb  
collisions at
 given impact parameter it generates charmonium states, which
are filtered through nuclear absorption, and then put in a hot de-confined
medium. The survival probability in it is  
$e^{-\Gamma\,t_p}$ where $t_p$ is the time spent in
this medium (see fig.\ref{suppression}(a))
and $\Gamma$ is the ``ionization'' rate. $\Gamma$  
is chosen to match the observed 
suppression for central events.To calculate the time inside plasma we need to distribute
these events in coordinate and momentum space.
The distribution in the transverse plane is given by Glauber theory,
like in
\cite{Kharzeev} and the distribution in the longitudinal direction is
presumed 
uniform.

 The question is then whether this ``semi-realistic''
 model provides good description of NA50 data
at all b, especially near the discontinuity,
The result of the calculation is shown in
 in Fig.$\,$\ref{suppression}(b). 
Because
the usual Glauber-type absorption on nucleons describes well 
the p-A and light lighter ion data (not shown) as well as
 peripheral PbPb data,
  we  take it out of the picture,  plotting
only 
 $deviations$ from it. As one can see from
 Fig.$\,$\ref{suppression}(b), 
the increased QGP lifetime
leads to additional suppression for intermediate $E_t=50-100 GeV$. 
 ``Leakage'' and other smearing effects do not
allow to reproduce a  jump seen in data,
as expected. However,  extra QGP lifetime does improve
agreement with most\footnote{The origin of a (non-statistical?) fluctuation
at $E_t\approx 100 GeV$ remains unanswered experimentally.} 
of the data points for $E_t>60 GeV$.
In particularly, it reproduces vanishingly small 
slope of the suppression in this region.

Potentially important way to separate different suppression
mechanisms is related with $p_t$-dependence of the suppression. 
We have therefore calculated $<p_T^{2}>_{J/\psi}$ vs. $E_T$, see  
Fig.$\,$\ref{suppression}(c).
The usual initial state parton re-scattering
leads to $<p_T^2>_{J/\psi}$ growing with $E_t$. However,
if  $J/\psi$ produced in the central region are destroyed
and only peripheral ones survive, then (as
suggested by Kharzeev et al)  this dependence  becomes inverted,
with a decrease at most central collisions. 
Our scenario leads to intermediate flat dependence,
because some $J/\psi$ from the center may still ``leak'' out. It was
encouraging to see that
  NA50 data (first shown at this conference after my talk) show exactly
such dependence!

  Finally, in \cite{ST_97} we have also 
 discussed  
the conversion of $J/\psi$ into its spin partner $\eta_c$.
This channel is one more potential ``sink'' for  
$J/\psi$ and to our knowledge has not been investigated previously: however our estimates of the rate of conversion show that it hardly can be essential.

\section{EVENT-BY-EVENT FLUCTUATIONS AND THERMODYNAMICS}

 Event-by-event fluctuations are  deviations of mean value
of some observable calculated in  an event,
relative to that averaged over the whole ensemble of events. 
  In general,  those
  can be of $dynamical$ or of
$statistical$ nature, and  I  focus on the latter
ones
(which are guaranteed to be there).
 Their sensitivity  to some non-trivial details of the
dynamics were first pointed out in  \cite{GM_92}. It was
 noticed that dispersion of the
   $<p_t>_{event}$ distribution strongly depends on
 a cascade model used. In particular, those based on
 ``initial rescatterings'' or 
  superposition of independent
pp events  predicted much larger fluctuations, compared to 
  models  with multiple re-scattering of secondaries. 
 This paper has triggered experimental studies (reported
at QM97 by Roland):  the  fluctuations observed by
the NA49 experiment were found to be
perfect Gaussian for several orders of magnitude, without any unusual tails.
 Its width
 is  very small, which clearly   rules out
any model of the former type.

  The essence of it is that the total $entropy$ generated in
the collisions shows up in the magnitude of fluctuations.
 In general, any statistical fluctuations can be derived from
the famous Boltzmann expression relating entropy S and probability P,
written in the Einstein form
$ P\sim exp(S) $.

  Before we turn to specifics, few general comments
 are appropriate. Applying thermodynamical theory of fluctuations
 we rely on 
   statistical independence of  different volume elements.
The fluctuations of temperature or
  particle composition should
be $independent$ on
whether all elements of the excited system have their freeze-out
at the same time or not, whether 
they freeze-out  at rest  in the same coordinate frame or
are all moving with different velocities.
The  relevant thing  is $how$
 the global entropy of the system S depends on the particular
observable under consideration.

  In general,
 the fluctuations accurately predicted by this expression neither
should   be   small, nor the
discussed system should have huge number of degrees of freedom.
The  requirement is that the system is equilibrated,  equally
populating all its available phase space\footnote{ As an 
example of such approach 
to  multi-hadron production reaction, let me
refer to two (my own  old) papers  \cite{Shu_oldthermo} where
such approach was used for the $K/\pi$ ratio dependence on
  pion multiplicity, as well as
probabilities of  exclusive channels in low energy
$\bar p p$ annihilation. Remarkably, simple ideal  gas formulae
for the entropy 
correctly predicted them,
 starting from reactions with  only 4 secondaries!}.

    The first example we consider \cite{Stodolsky,Shu_fluct}
is the  fluctuations in apparent\footnote{As discussed above,
the $m_t$ slopes are affected
by collective flow: but estimates show that the corresponding
correction fluctuate less and mostly cancels in ratios, especially for pions. 
}
freeze-out temperature $\tilde T$.
 Standard thermodynamics  tells us that temperature fluctuations are
 given by
$ P\sim exp[-{C_v(T)\over 2}(\Delta T/T)^2] $
where  $C_v(T)$  is the heat capacity of a hadronic matter.
It is an extensive quantity $C_v=T{\partial S\over
\partial T}|_{T,V}$, proportional to the total volume (or the
number of particles N) of the system, so the relative
 fluctuations  are $O(1/N^{1/2}$ as expected. 

  The key point  \cite{Shu_fluct} is that  $C_v(T)$
 has strong T-dependence, which can be used as
a ``thermometer".
 In the vicinity of the QCD phase transition  $C_v(T)$
has a peak. So, 
 if 
the observed hadrons are emitted directly 
from the QGP clusters/mixed phase (as suggest by some people
based on chemical freeze-out), 
the   fluctuations of $\tilde T$ should be very small.
  However, if  chemical and thermal
  freeze-outs are different
(as we strongly advocated above) and the observed
fluctuations  happen  in  cool gas, its 
 $C_v(T)$ is smaller.  We argued above that
   larger systems cool
 to lower $T_f$. Therefore we  end up with 
a (counter-intuitive) prediction:  (properly scaled)
 fluctuations for (central) heavy ion collisions should show  {\it stronger}
  fluctuations compared to medium ions or peripheral collisions.

  The second example  \cite{Shu_fluct} deals with
 fluctuations of 
occupation of 
particular bins in the histograms.
Deviations from trivial Poisson statistics are induced by 
 quantum statistics. For the ideal 
Bose-Einstein (Fermi-Dirac) gas one gets
$ <{\Delta n_k^2}>=n_k(1 \pm  n_k) $.
So, by measuring fluctuations in occupation
number of different bins one can measure the {\it quantum
  degeneracy} $n_k$
 of the
gas at freeze-out\footnote{The same quantity is in essence measured in 
the usual HBT: the difference is that we propose to detect
Bose-Einstein effect of $all$ pions in a bin, rather than a particular pair.}
  Let us briefly mention the magnitude of the effect.
 Pions in a chemically equilibrated gas
($\mu_\pi=0$) with zero momenta should have 
 fluctuations  enhanced  by a factor 1.5.
  If pions
  have a non-zero 
 $\mu_\pi$ (as advocated above), the effect becomes stronger: at $p_t=0$
Bose enhancement  reaches a factor of 2. 
  Finally,  if there are dynamical low momentum excess pions
(e.g. due to DCC)
one should  observe even larger fluctuations in the corresponding
bins.

\section{Summary and outlook}
  
  The main result of our  studies of the radial flow 
in central collisions 
 can  be understood in hydro framework only if the special care
is taken about the freeze-out. When it is properly done, all puzzles
are gone and the ``standard'' lattice-based
 EOS {\it including softness due to the QCD phase transition} 
approximately describe
the data, for all ions, both at AGS and SPS. Further work, 
in combination with elliptic flow data, is however needed to tell if this
description
is  unique.


One significant conclusion
is  existence of a rather large gap between
``chemical'' and ``thermal''  freeze-outs. For central
PbPb collisions at SPS the former corresponds to $T_{ch}\approx 160 MeV$
while the latter (in the very center) cools down to  $T_{ch}\approx 100-120 MeV$.
If so, the non-zero chemical potential for pions should appear:
there are first indication for reality of this efect in NA44 data.
Completely new set of ideas has recently emerged, relating  
event-by-event fluctuations and thermodynamics. Potentially it is very
useful way to learn  more about the freeze-out conditions. It is also
relatively simple tool for large-acceptance detectors like NA49
or STARS. 

  Finally, the non-trivial EOS leads to non-monotonous energy/impact
parameter dependence
of the dense matter lifetime, with the maximum corresponding to the
``softest point''
 \cite{HS_96}. We have speculated above that this effect contributes
 to $J/\psi$ suppression, and suggested a particular model to be
 compared with data. We hope that all experiments will now study the $E_t$
dependence of their observables, or  run at another 
(lower)
E/A. 

\end{document}